\documentclass[12pt]{article}

\usepackage{hyperref}
\usepackage{physics}
\usepackage{times}
\usepackage{pdflscape}
\usepackage{cancel}
\usepackage{graphicx}
\usepackage{xcolor}
\usepackage{soul}
\usepackage{caption}
\usepackage{dsfont}

\newenvironment{natabstract}{%
\begin{quote} \bf}
{\end{quote}}

\title{Emergent topological fields and relativistic phonons within the thermoelectricity in topological insulators}

\author
{Daniel Fa\'ilde,$^{1\ast}$ Daniel Baldomir$^{1\ast}$\\
\\
\normalsize{$^{1}$Departamento de Física Aplicada, Instituto de Investigacións Tecnolóxicas},\\
\normalsize{Universidade de Santiago de Compostela,}\\
\normalsize{E-15782 Campus Vida s/n, Santiago de Compostela, Spain}\\
\\
\normalsize{daniel.failde.balea@rai.usc.es; daniel.baldomir@usc.es}
}

\date{}

\begin{document} 
\captionsetup[figure]{labelfont={bf},labelformat={default},labelsep=period,name={Fig.}}

\baselineskip24pt


\maketitle


\begin{natabstract}
Topological edge states are predicted to be responsible for the high efficient thermoelectric response of topological insulators, currently the best thermoelectric materials. However, to explain their figure of merit the coexistence  of topological electrons, entropy and phonons can not be considered independently. In a background that puts together electrodynamics and topology, through an expression for the topological intrinsic field, we treat relativistic phonons within the topological surface showing their ability to modulate the Berry curvature of the bands and then playing a fundamental role in the thermoelectric effect. Finally, we show how the topological insulators under such relativistic thermal excitations keep time reversal symmetry allowing the observation of high figures of merit at high temperatures. The emergence of this new intrinsic topological field and other constraints are suitable to have experimental consequences opening new possibilities of improving the efficiency of this topological effect for their based technology.
\end{natabstract}

The purely experimental fact that nowadays the best room temperature thermoelectrics are topological insulators (TIs) \cite{Xu2017}, not only deserves a deep explanation to satisfy our scientific curiosity, but it also could greatly help in the development of new renewable energy technologies. The efficiency of heat transformation into electricity (or vice versa) in a thermoelectric material is measured with a parameter called the dimensionless figure of merit $ZT$, where $Z$ is the figure of merit and $T$ the absolute temperature. Its highest value is around 2,4 at 300K for Bi$_2$Te$_3$/Sb$_2$Te$_3$ superlattices \cite{Venkatasubramanian2001}. These structures are built with topological insulators for decreasing the thermal conductivity as much as possible and take advance of the highly conducting edge channels present in TIs to provide a high efficient thermoelectric response \cite{Bernevig1757,Baldomir2019}. Nevertheless, the technological applications still need a higher value to the previously mentioned \cite{LIU2015357}. Besides the experimental difficulties to obtain high-quality thin-films, eliminate bulk carrier transport or adjust the Fermi level to lie in the small topological gap there is not a physical model to guide us towards a new generation of thermoelectric devices by treating phonons or oscillations produced by temperature in the topological context. The problem is that physics is fundamentally local and topology must enter through those tools. With this purpose, we need to connect topology and electrodynamics in order to introduce relativistic phonons and thermal excitations together with the topological electrons at the surface. The scenario which arises after this study, supported on an obtained expression for the topological intrinsic field, solves most of the difficulties involved in the topological thermoelectricity and allows to find experimental work to confirm it.

The global characteristics of TIs are printed on their band structure within their first Brillouin zone (BZ), concretely through their curvatures and Berry's phases. These materials are both  insulators in the bulk and semimetals on their surface where they contain Dirac points on account of the band crossings. But the bands are not the unique factor to take into account to determine the topology. One fundamental ingredient is the dimensions of the spacetime where the Hamiltonian is defined. That is why generally, for a given Hamiltonian $H(k)$ following Schr\"{o}dinger equation, we are able to define a map from the BZ to the manifold of their symmetries $\nonumber X(k): BZ \longmapsto \frac{U(n)}{U(n-i)\times U(i)}$ by means of a unitary matrix $U(n)$ which diagonalizes the Hamiltonian for $n$ states (Bloch bands) being $i$ the occupied ones and $n-i$ the unoccupied with respect to the Fermi level. This allows calculating the number of maps $X(k)$ which cannot be deformed continuously each other using the homotopy group obtaining $\pi_1[X(k)]=\pi_3[X(k)]=0$ and $\pi_2[X(k)]=\mathds{Z}$, being $\mathds{Z}$ an integer and each subindex the dimension of the spacetime \cite{Nakahara}. That means that there is only one non-trivial topology associated with the direct relabeling of the bands in two dimensions. Or in other words, band insulators with different integers cannot be continuously deformed into each other without crossing a quantum phase transition. This is what happens, for instance, in the Integer Quantum Hall Effect, where the number of chiral edge states are the integers associated the above homotopy group \cite{PhysRevLett.95.146802,PhysRevB.78.195125,PhysRevB.79.195322,PhysRevLett.51.51}. But in one or three dimensions we need to introduce more symmetries of the Hamiltonian if we want to take into account their non-trivial topology. In TIs this symmetry is the time-reversal symmetry $\hat{T}$ which induces a Kramer's degeneracy \cite{PhysRevLett.95.226801,PhysRevLett.97.236805}, such that the square of $\hat{T}$ operator is equal to $-1$ for half-spin electrons (fermions), acting on the map  previously defined as $\hat{T}X(k)\hat{T}^{-1}=X(-k)$. This enables us to introduce a new topological number by the discrete cyclic group of two elements $Z_2$ called spin-Chern number \cite{PhysRevLett.95.226801,PhysRevLett.97.036808}. The form to do it is somewhat subtle dividing the Hilbert space into two parts, one for each kind of Kramers states, and calculating the Chern number on them for extending the definition of the topological index \cite{RevModPhys.82.3045,PhysRevB.78.195424}.

Once we know how the non-trivial topology can be determined on the bands, we need to figure up how the electromagnetic fields and heat exchange behaves physically under such non-trivial topologies. For such aim, we need to employ the axion electrodynamics which enlarges the action with the two Lorentz invariants associated to the fields. That is to say, to add the pseudo scalar quantity $\frac{1}{2}\epsilon_{\mu\nu\alpha\beta}F^{\mu\nu}F^{\alpha\beta}$, besides the usual scalar $F_{\alpha\beta}F^{\alpha\beta}$ which is enough to provide Maxwell electrodynamics in the trivial topological vacuum. Then, we have the axion action $\mathcal{S}=\int dx^4(\frac{1}{4}F_{\alpha\beta}F^{\alpha\beta}-\frac{e^2}{32\pi^2\hbar}\theta(r,t)\epsilon_{\mu\nu\alpha\beta}F^{\mu\nu}F^{\alpha\beta}+A_{\mu}J^{\mu})$ which leads directly to the equivalent Maxwell equations in TIs, $\boldsymbol{\nabla}(\boldsymbol{E}+2{\alpha c} (\frac{\theta}{2\pi})\boldsymbol{B})=\frac{\rho}{ \epsilon_0}$ substituting the Gauss law and $\boldsymbol{\nabla} \times \boldsymbol{B}=\mu_0\boldsymbol{J}+\frac{1}{c^2}\frac{\partial \boldsymbol{E}}{\partial t}+\frac{2\alpha}{c}[\boldsymbol{B}\frac{\partial}{\partial t}(\frac{\theta}{2\pi})+\boldsymbol{\nabla}(\frac{\theta}{2\pi})\times \boldsymbol{E}]$ instead of the Maxwell-Ampere one, being $\alpha= \frac{e^2}{\hbar c}$ the dimensionless fine structure constant in electrostatic cgs units, where the electric charge is given in statcoulombs. The other two equations, Faraday and non-existence of isolated magnetic poles, maintain the same form \cite{Wu1124,PhysRevLett.102.146805,PhysRevB.95.075137}. It is necessary to remark that the phase $\theta(r,t)$ needs to be a function of the space and time if the new electrodynamics wants to take into account the topological background. Actually, in the TI the proper choice is $\theta(r,t)=\frac{e}{\hbar}\Lambda(r,t)$, being $\Lambda(r,t)=\frac{\hbar 2\pi r}{ea}$ a gauge function associated to the degree of freedom of the electromagnetic potentials where $a$ is a lattice constant and $|r|$ de modulus of the distance given in the electromagnetic fields. These ingredients are enough to connect $\theta$ with the non-trivial topologies of the lattice using Gauss-Bonnet, however, not with the bands through the Chern number. To reach such result the translation of the topological information into a physical field, named as $b$, is necessary first. This will be done through the Berry curvature $\Omega_{k_x,k_y}=-2Im \bra{\partial_{k_x}n}\ket{\partial_{k_y}n}$ defined on the non-trivial bands of a 2D Dirac Hamiltonian which behaves as a spin-dependent magnetic field in the $k$-space and whose integral determines the Chern number $C$.

Supported by this topological field, which is intrinsic to these materials and consistent with their special electromagnetic background, we are able to introduce and interpret the thermodynamic part associated to the phonons in TIs (Fig.1). This is done by including oscillations in the Dirac Hamiltonian, i.e., relativistic phonons associated to the Dirac oscillator \cite{moshinsky1989dirac}. By means of the adiabatic mechanism, we are going to establish an equivalence between the phonon field $\boldsymbol{\omega}$ of the Dirac oscillator and the field $\boldsymbol{b}$ containing the information of the dynamic and the robustness of the topological regime. This is a crucial result for the topological formalism of the thermoelectricity. On the one hand, we demonstrate in a direct way how relativistic phonons enters into the topological context, modifying the Berry curvature and the axion $\theta$ angle on the lattice allowing heat-electricity transformation. On the other hand, we give an explanation of why topology is preserved at high temperatures in most of the compounds which exhibit it. The previously mentioned relationship between the field $\boldsymbol{b}$, which easily reaches values in the order of Teslas for each spin subsystem with the typical parameters in 3DTI thin films, and $\omega$ defines a limit ($>$THz) for the frequencies tolerated by the system without involving entropy change. Under these conditions, we would find a temperature regime in which such excitations would not break the quantum coherence necessary for the conservation of the topological signatures allowing  the observation of the high figures of merit associated to the topological states around room temperature.

\section*{Results}
The family of 3DTIs Bi$_2$Se$_3$, Bi$_2$Te$_3$, Sb$_2$Te$_3$  has a special interest by the fact of being topological besides including on their members the most efficient thermoelectric material up to now. The highly conducting edge states, provided by the topology, are predicted to be responsible for their high figure of merit in low thermal conductivity conditions \cite{Baldomir2019,PhysRevB.81.161302}. However, this must not be the unique ingredient to explain the thermoelectricity in these materials where the coexistence of time-reversal symmetry and non-zero temperatures, which usually involves entropy change, might cause a conflict. Given that, our starting point must be a 2D effective Dirac Hamiltonian used to describe the physics inside 2DTIs as well as in 3DTIs thin films, i.e. when the thickness of a 3DTI is enough small to overlap its top and bottom surface states forcing them to be placed at the edge \cite{PhysRevLett.101.246807,PhysRevB.81.115407,Zhao2014}. 

\begin{equation}
H_{2D}({\bf k})=\left[\begin{array}{cc} {H_+} & {0} \\
0 & H_- \\
\end{array} \right], \qquad 
\qquad H_{\pm} = 
\left(\begin{array}{cc} \pm M({\bf k}) & \hbar v_F k_-\\ \hbar v_F k_+ & \mp M({\bf k}) \end{array}\right) \\
\label{2DHamiltonian}
\end{equation}
Here $k_\pm=k_x\pm ik_y$, $M({\bf k})=M-\mathcal{B}k^2$ is the effective mass in the solid, $k^2=k_x^2+k_y^2$, $v_F$ is the Fermi velocity, $\hbar$ is the Planck's bar constant and the basis has been rearranged to be $\left[ \psi_{1\uparrow}, \psi_{2\downarrow}, \psi_{2\uparrow}, \psi_{1\downarrow} \right]$, allowing the separation of the Hamiltonian into two non-interacting time-reversal counterparts $H_\pm$ which can be treated independently \cite{PhysRevB.81.115407,PhysRevB.82.165104}. The energy spectrum of equation \eqref{2DHamiltonian} define two non-interacting Dirac hyperbolas centered at the $\Gamma$ point for which conduction and valence bands for $H_+$ have an associated  Berry curvature

\begin{equation}
    \boldsymbol{\Omega}^{c}_{k_xk_y}=-\boldsymbol{\Omega}^{v}_{k_xk_y}=-\frac{\hbar^2v_F^2(M+\mathcal{B}k^2)}{2[(M-\mathcal{B}k^2)^2+\hbar^2v_F^2k^2]^{3/2}} \boldsymbol{\hat{z}} 
\label{curvature}
\end{equation}
which is spin and band dependent, resulting in the opposite sign for $H_-$ ($M\rightarrow-M$). This Berry curvature defines the Chern number $C=1/(2\pi) \int \boldsymbol{\Omega}d \boldsymbol{k}$ which in the non-trivial regime of $H_{2D}$, given by the condition $M\mathcal{B}>0$, is an integer equal to $\pm1$ which also shares the same dependence of $\boldsymbol{\Omega}$ and its responsible of transport quantization \cite{PhysRevB.81.115407}. Essentially, the Berry curvature plays the role of a magnetic field in the k-space obtained through the rotational of the Berry potential $\boldsymbol{A}$. This allows us to consider in a TI, the presence of a pair spin-momentum locking orbits associated to the topology, which present a quantized flux ($C \; h/e$) in terms of the Chern number $C$ and whose sum obviously gives zero due to $\hat{T}$ symmetry (Fig.2). Of course, this is consistent with the fact of why in the presence of an in-plane electric field we can talk about opposite transverse spin currents which in the edge produce a quantized electrical conductance $G=(C_+-C_-) e^2/h$, being $C_\pm$ the Chern numbers associated with the branches $H_\pm$ \cite{PhysRevLett.96.106802,Konig766}. But, as in all the parts of the physics, the phenomenology must be associated to a physical field, which in this particular case, must contain the topology. The answer lies in the translation of the Berry curvature into the real space. This can be done by noticing that under small gap conditions, as it happens for 3DTIs in the thin-film limit, the non-trivial Berry curvature $\boldsymbol{\Omega}_{k_x,k_y}$ associated to the states of equation \eqref{2DHamiltonian} has the form of a single peak Gaussian-like function centered at the $\Gamma$ point \cite{PhysRevB.81.115407}, which has characteristic length small enough to consider an equivalent magnetic field $\boldsymbol{b}$ constant along the bulk crystal and whose magnitude must be determined by the constraint that its flux is quantized and equal to $h/e \; C$, i.e., $\hbar/e \int \boldsymbol{\Omega}_{k_x,k_y} d\boldsymbol{k} = \int \boldsymbol{b} d\boldsymbol{S}$. 

Given that $b$ can be extracted from the integral, we need now to estimate the area defined by the topological electrons on their motion. This surface element $\Delta \boldsymbol{S}$ can be obtained in an original way by applying Heisenberg's uncertainty principle matching the quantum conductance ($e^2/h$) with the conductivity $\sigma=\Delta S^{-1} \frac{e^2 \tau}{m_{e}}$ in the Heisenberg limit ($\tau=\hbar/\Delta \varepsilon$) being $e$ the electric charge, $\tau$ the scattering time and $\Delta \varepsilon=2\xi$ the energy uncertainty which can be considered to be in the order of the energy gap given the low energy nature of the topological electrons \cite{Batra2002}. In this way, we obtain the following expression for the field $\boldsymbol{b}$
 
 \begin{equation}
    \textbf{b}=\frac{2 m_e \xi}{\hbar e} C \; \hat{\textbf{z}} \approx \frac{2m_e^2v_F^2}{\hbar e} C \; \hat{\textbf{z}}
\label{field}
\end{equation}
where the electron effective mass can be considered as $m_{e}=M/v_F^2$ neglecting any contribution from the Hamiltonian parameter $\mathcal{B}$, that gives us the information about the localization or delocalization of the bands in the space, and limiting to materials that present $v_F^2>>2M\mathcal{B}/\hbar^2$. This approximation, easily fulfilled thanks to the small gap and high Fermi velocity that typically characterizes 3DTI thin films ($M\approx -25$ meV, $v_F=6.17 \; 10^5$m/s), determines for these values an equivalent field for electrons on the surface $\abs{b}\approx5$T consistent with the robustness that characterizes topological surface states and whose sign, determined by the Chern number, keeps time-reversal symmetry intact. Note that in equation \eqref{field} $m_e$ must not change its sign when we pass from $H_+$ to $H_-$ ($C\rightarrow - C$) given that its inversion has already been considered  in the conductivity change $e^2/h\rightarrow -e^2/h$. So, our formalism defines an effective area $\Delta S=h\hbar/(2m_e \xi)$ for the electrons inside TIs and a magnetic field $b$ with opposite sign for each branch $H_\pm$, which attending to the full Hamiltonian, results in a special spin-dependent interaction consistent with the singular dynamic of the topological regime and the band inversion. The obtained expression for the field $b$ also shares a direct correspondence with the critical magnetic field $B_c=m^2c^2/(\hbar e)$ necessary for two photons to create Schwinger pairs in the vacuum \cite{PhysRevD.72.105004}, adapted to the particular context of the TIs due to the small gap and the substitution of $c$ by $v_F$, making this quantity experimentally accessible.

As we are going to see, now we are in a position to interpret the role of relativistic oscillations in TIs. It is straightforward to show that the abstract remarks defined above are closely related to the phenomenology of phonons in the Dirac oscillator Hamiltonian $i\hbar (\partial \psi / \partial t)=[v_F \boldsymbol{\alpha}(\boldsymbol{p}-im \boldsymbol{r}\omega\beta)+mv_F^2\beta] \psi$, which also preserves time-reversal symmetry and incorporates a linear correction in $r$ to the electron momentum $p$ in the form of a magnetic field $B=2m\omega/e$. In a 2D Hamiltonian as equation \eqref{2DHamiltonian} the perturbation introduced in the Dirac oscillator enters in the same way of a magnetic field $B$ in the $z$-direction with opposite sign (just like $b$) for each branch $H_\pm$  with the usual substitution $2\omega=eB/m$ with a difference of a factor 2 that comes from the non-minimal coupling present in the Dirac oscillator equation to guarantee of having a harmonic oscillator in its non-relativistic limit \cite{moshinsky1989dirac,Andrade_2014}. Precisely in this limit, is where a spin-orbit coupling of strength $2\omega/\hbar$ arises motivated by the spin-dependent magnetic interaction introduced in the system. The Dirac oscillator (Supplementary Information) has been analyzed in different studies \cite{moshinsky1989dirac,PhysRevA.76.041801,Andrade_2014,Rozmej_1999}, however, it has not been treated into the adiabatic formalism. Only then, we are going to be capable of visualizing the role of phonons inside the topological context and finally into the mechanism of thermoelectricity. Given that the Hilbert space was divided into two time-reversal counterparts, we are able to work in only one of the subsystems $H_\pm$ where the adiabatic correction to the energy eigenstates can be calculated \cite{shen2012topological,vanderbilt2018berry}. Lorentz force provided by $\boldsymbol{\omega}$ increases or decreases (depending on its sign) the electron momentum in the direction $i$ proportionally to their perpendicular components. In that way, we can formulate the temporal variation of the momentum as $\partial k_i / \partial t=\epsilon_{ijk} k_j \omega_k$ and, regrouping terms, write the correction to the energy states given by the Dirac oscillator 

\begin{equation}
    \ket{n} \rightarrow \ket{n} + \frac{\hbar \omega}{4\xi^2} \hbar v_F k \ket{m}
\end{equation}
where $\ket{n}$ and $\ket{m}$ are the positive and negative energy eigenstates of $H_+$. From here, we compute after some tedious algebra the corrections to the Berry curvature of the bands $\Omega^{i}_{k_xk_y}=-2Im \bra{\partial_{k_x} i}\ket{\partial_{k_y}i}$. Two cases were analyzed, both sharing an ability to modulate the Berry curvature (Supplementary Information). The first one, considering a purely uniform $\omega$, represents the phonons or oscillations with a constant energy dispersion and their corrections have the form of a function similar to the Berry curvature which changes its sign at some point $k$ (Supplementary Fig. S1). Besides being quite restrictive, these modes will not be suitable to introduce in the thermoelectric mechanism. The second case considers an explicit energy dependence on the phonon frequency, i.e. $\hbar \omega=\lambda \xi$, which not only could maximize the coupling with the topological electrons given its relativistic energy dispersion but at the same time is consistent with its apparent relation with the intrinsic topological field $b=2m\omega/e$, with no more ingredients that substituting equation \eqref{field} into the previous relation. If we are in the right way, corrections will give back to a function of the type of the Berry curvature given that $b$ comes from its translation onto the real space. Thus, made the calculations, curvature corrections turn out to be

\begin{equation}
    \Omega^n(k) \rightarrow \Omega^n(k) - \lambda\frac{M}{\xi} \Omega^n(k)
\end{equation}
where $\lambda$ ($\in [0,1]$) is a dimensionless parameter measuring the relative strength of $\omega$ with respect to $b$. The obtained results, plotted in Fig.3, demonstrates how phonons and oscillations can be introduced into the context of TIs modulating the Berry curvature and hence the field $b$ even when the perturbation is not small compared with the energy of the system but whose variation can be adiabatic. For the limit case $\lambda=1$, i.e. $\omega=eb/2m$, the obtained first-order correction have the same height at $\Gamma=0$ than the unperturbed Berry curvature. This confirms the correctness of the expression for the field $b$ and its interpretation as a measure of the topological robustness and as a critical value for the strength of in-plane oscillations and external forces (strain, spin-orbit) supported by the surface states. Beyond this limit, the Berry curvature could change its sign and force a topological phase transition to the trivial regime. Of course, these effects are not produced by every phonon mode presented in the crystal but some specific phonons for which the concept helicity must be involved in order to couple topological electrons and capable to cause a correction to the momentum of $\partial k_i / \partial t=\epsilon_{ijk} k_j \omega_k$, i.e., relativistic. This constraint, that will need a suitable phonon dispersion, has been shown to be on the energy range needed for polar optical modes in Bi$_2$Te$_3$ and Bi$_2$Se$_3$ \cite{Heid2017}. The expressions derived above are also valid in the case of taking into account the dependence of the mass term $M(k)=M-\mathcal{B}k^2$ on the Hamiltonian parameter $\mathcal{B}$ (Supplementary Fig. S2). In all cases, as it is expected, it can be show that the corrections introduced respects the intrinsic particle-hole $\hat{C}$ and $\hat{T}$ symmetries of $H_{2D}$.

Both, electrodynamics and bands, have a common degree of freedom given by the gauge transformations within the Abelian group U(1). Mathematically they also share non-trivial topological features where the homotopy group $\pi_1(U(1))$ is equivalent to the integer numbers: the bands through the Berry curvature and the electrodynamics through torus constraints with genus equal one. Let us to develop it deeper from the physical consequences. Using the periodicity of the Brillouin zones, we can define the function $\Lambda(r,t)=\frac{\hbar 2\pi r}{ea}$, where a is the lattice constant, which allows a transformation of the potentials such that $A_{\mu}\longrightarrow A'_{\mu} =A_{\mu}+\partial_{\mu} \Lambda(r,t)$. This makes the space-time to be on a torus of four dimensions $T^4=T^2 \times T^2$ for the electromagnetic fields and taking $\pi$ degrees between $E$ and $B$, i.e. sharing a spatial direction. This leads to a quantized electric and magnetic fields, $B=n \frac{h}{e a^2}$ and $E=n' \frac{h c}{e a^2}$, where $n,n' \in  \mathcal{N}$ are natural numbers which determine the topological sector. Coming back to the $EB$ term of the action we can see how this part of the action is quantized making $\theta$ belong to the interval $[0,2\pi)$ and therefore obtaining a topological vacuum angle that usually is written as $\ket{\theta}=\sum_\nu exp(i \nu\theta)  \ket{\nu}$, being $\nu$ the winding number \cite{PhysRevB.95.075137}. The defined electromagnetic background can be now easily connected with the topology of the bands through the intrinsic topological field $b$, substituting the area $a^2$ by the one $\Delta S=h\hbar/(2m_e \xi)$ obtained for the electrons in a TI, resulting that

\begin{equation}
    B=\frac{2m_e\xi}{\hbar e}\; C\equiv b
\end{equation}
where $n$ is now interpreted as the Chern number $C$. Notice that $b$ is associated to the curvature of the bands, while $B$ is a pure magnetic field of axion electrodynamics carrying information of the non-trivial topology on the spacetime. Both things are conceptually very different, one worked with Chern numbers and the other with genus using Gauss-Bonnet theorem; but physically they are connected linking the singularities of the bands and the real spacetime. Consistently, by making the same substitution we also find an expression for the electric field $E=2m_e^2v_F^3n'/\hbar e$ inside TIs, obtaining a value that determines the field needed to force a topological phase transition and which is again related with the corresponding critical field $E_c$ in the vacuum needed to create electron-hole pairs\cite{Baldomir2019,Pan2015,Heisenberg1936}. This is all that we can have for the electromagnetic fields within the TI with constant $\theta$, they are fixed and and hence can not intervene in the dynamics, but we can overcome this difficulty just introducing phonons that move the different topological sectors. In this case the lattice constant depends of time  and also the above electric and magnetic fields through the winding numbers $n$ and $n'$. Berry curvature also do it, as we shown, keeping gauge invariant the bands with respect to the electromagnetic potentials. 

Therefore, we can tackle the phenomena of topological thermoelectricity by means of the field $b$, equivalently to a treatment through a Chern-Simons action \cite{Baldomir2019}, that varies due to the effects of phonons or thermal oscillations.  The role of temperature into the topological electron transport can joined up now by assigning to the energy $\Delta\varepsilon$ an adiabatic temperature dependence $k_B T$. Note that there is an infinite number of orbits for which the relation between $b$ and $\Delta S$ defines a quantized flux $h/e$. Thus, when a phonon or a thermal excitation couples to a topological electron, the electron must move towards another orbit consistent with its new $b$ in order to maintain the quantum flux $h/e \; C$ constant. Consistently, in presence of an thermal gradient all the states share the same electrical conductance $G=I/V=e^2/h$, that can be obtained again through the Heisenberg relation used before, i.e. $\tau \Delta \varepsilon= \hbar$, being $V=2\pi \Delta \varepsilon/e$ and $I=e/\tau$ the electric potential and the electric current respectively. Thus, we can obtain the change in the electric potential due to thermal effects
\begin{equation}
    V= V' + \frac{2\pi}{e} C k_B T
    \label{electric potential}
\end{equation}
and also its associated electric field
\begin{equation}
    E= -\frac{2\pi}{e} C k_B \boldsymbol{\nabla} T - \frac{2\pi}{e} \frac{\partial C}{\partial r} k_B T \; \hat{\boldsymbol{r}}
    \label{electric field}
\end{equation}
The resulting expressions are identical to that obtained in ref. \cite{Baldomir2019} introducing the thermodynamic part in TIs through a Chern Simons action. The same result is deduced from $b$ providing the electric response in TIs not only for a thermal field in general but also for relativistic phonons. Thus, the second term in equation \eqref{electric field} makes reference for strong thermal perturbations which are able to change the Chern number of the system, producing an anomalous Seebeck contribution $S=\frac{2\pi}{e} k_B \frac{\partial C}{\partial T}$ associated to the creation of electron-hole Schwinger pairs. In contrast, the first one take into account the perturbations whose strength is below the limit defined by the topological intrinsic field $\boldsymbol{b}$ and as we shown before, they can produce changes in the Berry curvature without changing the Chern. Focusing on its regime, the explanation of why we can observe the high thermoelectric response associated to the topological edge states around room temperature is immediate. Consider for instance a coherent process (Fig. 4), in which a phonon or a thermal excitation couples to a topological electron increasing the field $b$ and producing an electric field $\boldsymbol{E}=-2\pi/e \; Ck_B \boldsymbol{\nabla}T$ associated to its new configuration. As it can be easily calculated, the Seebeck coefficient $S=\partial V/\partial T=2\pi/e \; k_B C$, and hence the entropy of the system, does not depend on temperature. That is, all the possible final states, associated to a small or a large adiabatic perturbation, share the same Seebeck coefficient, keeping the entropy constant, up to the limit in which the second term of equation \eqref{electric field} must be considered. So, even in the worst case, where perturbations go against the topological intrinsic field $b$ decreasing it, the coherence needed to observe topological signatures would still being conserved up to a temperature $T\sim 2mv_F^2/k_B$. Notice that the sum of all the entropy associated with these processes is zero, as the Chern numbers do when there is time-reversal symmetry. The translation of these results to edge physics is clear. The electric field generated due to the temporal variation of their intrinsic $b$ turns into an enhancement (coherent) or reduction (decoherent) of the relative moment between the two helical currents, however, the edge states remain to be ballistic with a quantized conductance $G=e^2/h \; C$ up to the Chern changes to zero and the coherence is lost.

In summary, we present a formalism based on a purely effective topological intrinsic field $b$ which allows to treat topology in a new way measuring its robustness at the same time it allows to incorporate relativistic phonons into the topological context. This is the first time that a given field or perturbation is demonstrated to modulate the Berry curvature of the bands preserving the intrinsic symmetries of the TIs. But moreover, this enable us to define a regime ($\omega<eb/2m$) in which oscillations are not enough large to make changes in the Chern number ($\partial C/\partial T=0$). In this situation, phonons and thermal excitations could couple to electrons without changing entropy, and then, the Seebeck coefficient $S$ and maintaining the coherence necessary to keep the quantum conductance $G=e^2/h \; C$ of its highly conducting channels invariant too. This fact is valid for a wide range of values given the magnitude of $b$, which for the parameters characterising 3DTI thin-films ($M\approx 0.025eV$, $v_F=6 \; 10^5$m/s) defines a field of $5$T and a frequency limit above the THz, both quantities generalizable to any topological insulators in 2D or for 3DTIs in thin film conditions with no more ingredients as their band gap, Fermi velocity and Chern number. This explains exactly how temperature and time-reversal symmetry coexist allowing the observation of the topological signatures and their associated high figure of merit at room temperature \cite{PhysRevLett.107.210501,Park2015,Zhao2014}. The ingredients provided are basic to derive the expression for the topological figure of merit ZT associated to the Kramers edge states in zero lattice thermal conductivity conditions \cite{Baldomir2019}, but additionally, we define in which circumstances the current experimental limit can be overtaked \cite{Venkatasubramanian2001}.

A new route to find high efficient thermoelectric devices by means of a special coupling between electrons and phonons at the surface is described \cite{PhysRevLett.98.166802}. In such situation there should be a depression in the number of phonon modes available in the system to contribute to thermal transport meaning a reduction in the lattice thermal conductivity and a heat-electricity transformation. This can be translated from an enhancement of electron-phonon coupling as it has been observed in 3DTIs Bi$_2$Se$_3$ and Bi$_2$Te$_3$ where polar optical phonon modes look to couple strongly with Dirac electrons when the Fermi level lies close to the Dirac point  \cite{Heid2017}. The map between these results and the model seems immediate since we are considering in-plane oscillations (polar) in our surface Hamiltonian at the same time as the enhance of electron-phonon coupling takes place for frequencies satisfying the condition $2\omega=eb/m$ previously underlined. This could imply an alternative method to decrease lattice thermal conductivity for certain phonon modes compared with other techniques (superlattices, impurities, dislocations) or lead to exotic phenomena \cite{Venkatasubramanian2001,tretiakov2010large,Chakraborty2017}. Additional  experimental scenarios can be proposed to support our results. Besides a relativistic electron-phonon coupling and a high thermoelectric response, the study of the topological intrinsic field $b$ can lead to more direct verification. Defined as an outcome of the topological robustness it enables us to define the critical magnetic and electric fields on each spin subsystem $H_\pm$ that could be measured.

\section*{Acknowledgements}
Authors acknowledge to CESGA, AEMAT ED431E 2018/08, PID2019-104150RB-I00 and the MAT2016-80762-R project for financial support.
\section*{Author contributions} 
D.F. and D.B. conceived the problem, made the calculations and wrote the manuscript. 
\section*{Competing Interests} The authors declare no competing interests.

\begin{figure}
    \centering
    \includegraphics[scale=0.90]{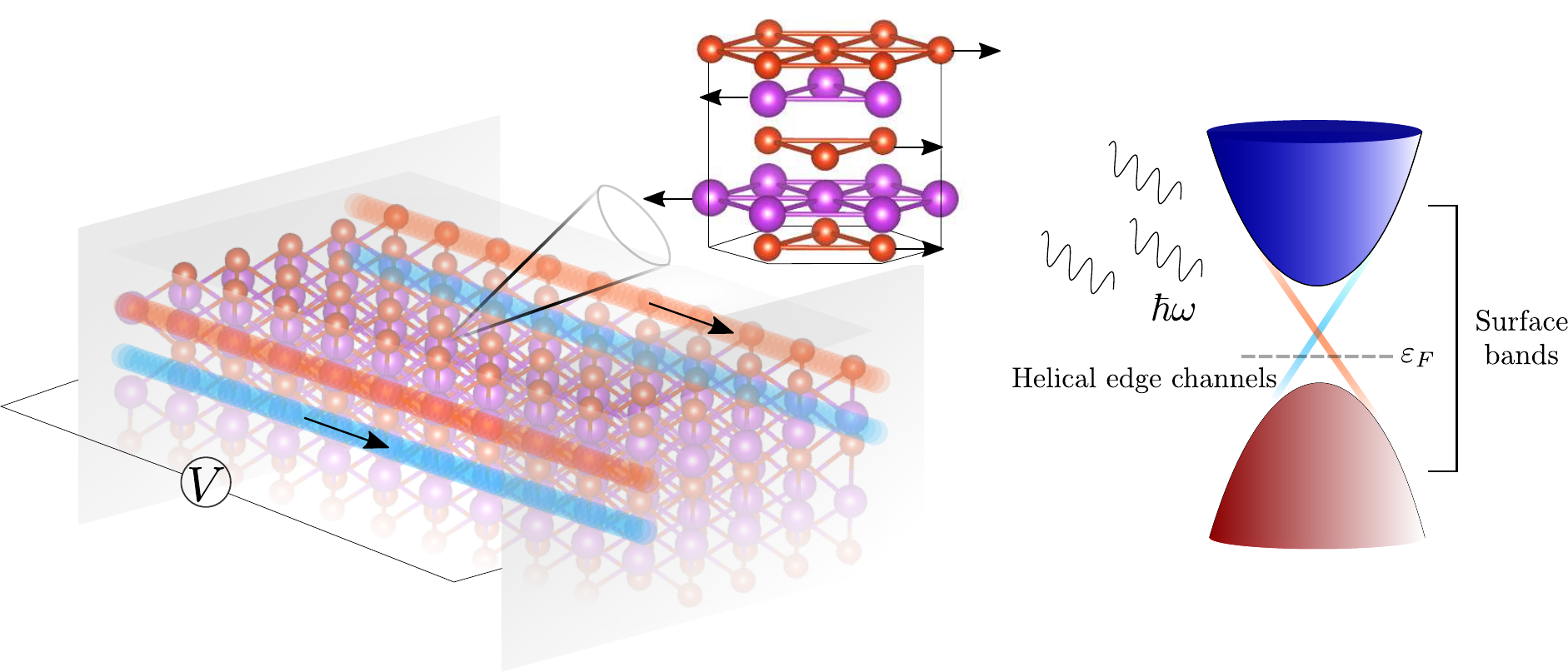}
    \caption{\textbf{Lattice oscillations in the topological electronic transport.} Schematic illustration of helical edge states and surface of a 3DTI in thin-film conditions. The energy spectrum is displayed on the right side of the panel. The quintuple layer describes the addition of in-plane lattice oscillations, in this case represented by a polar phonon mode, to the topological electronic transport.}
\end{figure}

\begin{figure}
    \centering
    \includegraphics[scale=0.50]{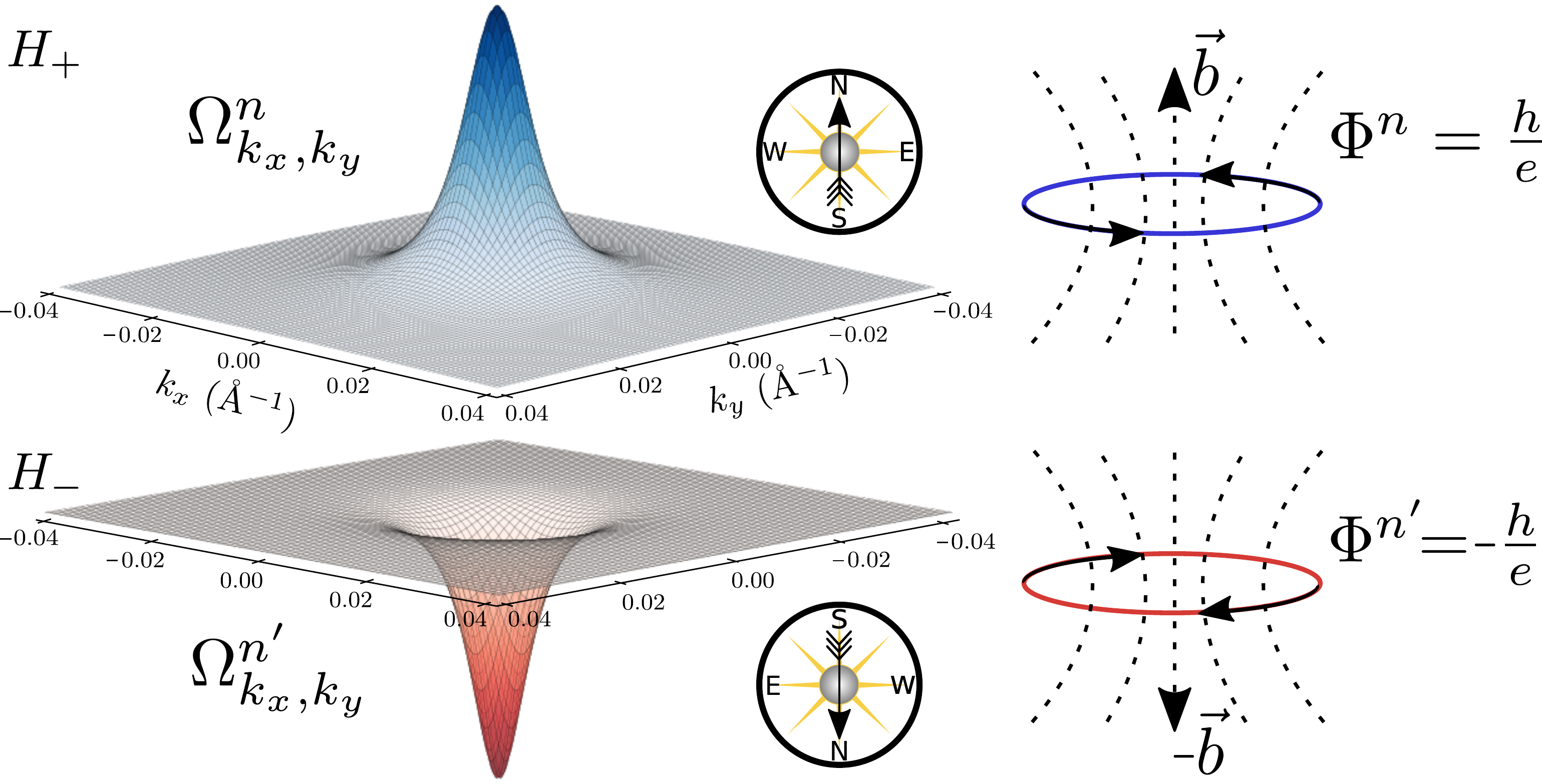}
    \caption{\textbf{Non-trivial Berry curvature and effective flux quantization in TIs.} Non-trivial ($M<0$, $B<0$) Berry curvature $\boldsymbol{\Omega}_{k_x,k_y}$ for the positive energy eigenstates of $H_\pm$ labelled as $\ket{n}$ and $\ket{n'}$ respectively. The compass indicates the orientation of the field  felt by the electrons on each band. In the bulk of a TI, this curvature allows considering the existence of helical orbits with an associated flux $\Phi=\hbar/e \; C$, being $C=\pm 1$ the Chern number associated to the conduction bands of $H_\pm$.}
    \label{Fig2}
\end{figure}

\begin{figure}
    \centering
    \includegraphics[scale=0.80]{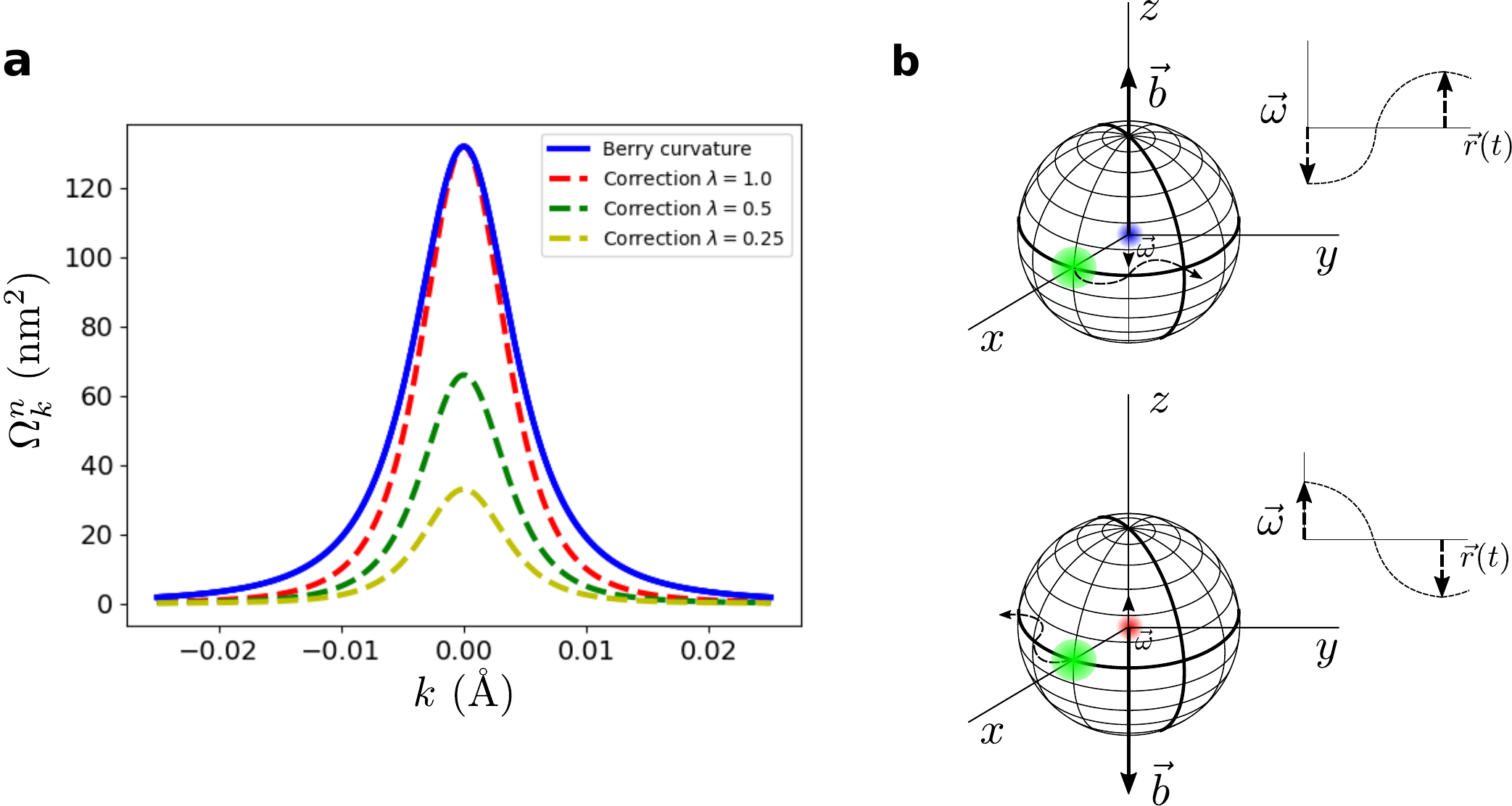}
    \caption{\textbf{Berry curvature correction and b field interpretation}. (\textbf{a}) Unperturbed Berry curvature (blue solid line) of the conduction band ($H_+$) of a topological insulator ($M<0$) and first-order correction to it (dashed lines) for different energy dependent frequencies below the critical frequency $\omega_c=eb/2m$. The parameters used are $M=-0.025$ eV, $B=0$ and $v_F=6.17$ $10^5$m/s. (\textbf{b}) Schematic translation of Berry curvature modulation from the electronic point of view (blue and red points) where in-plane nucleus  (green circles) displacements are showed to change the field $b$ defined on each orbital motion. Red and blue points represent electrons with opposite spins for which nucleus move in opposite directions attending to the helical nature of electronic motion in a TI.}
\end{figure}

\begin{figure}
    \centering
    \includegraphics[scale=0.9]{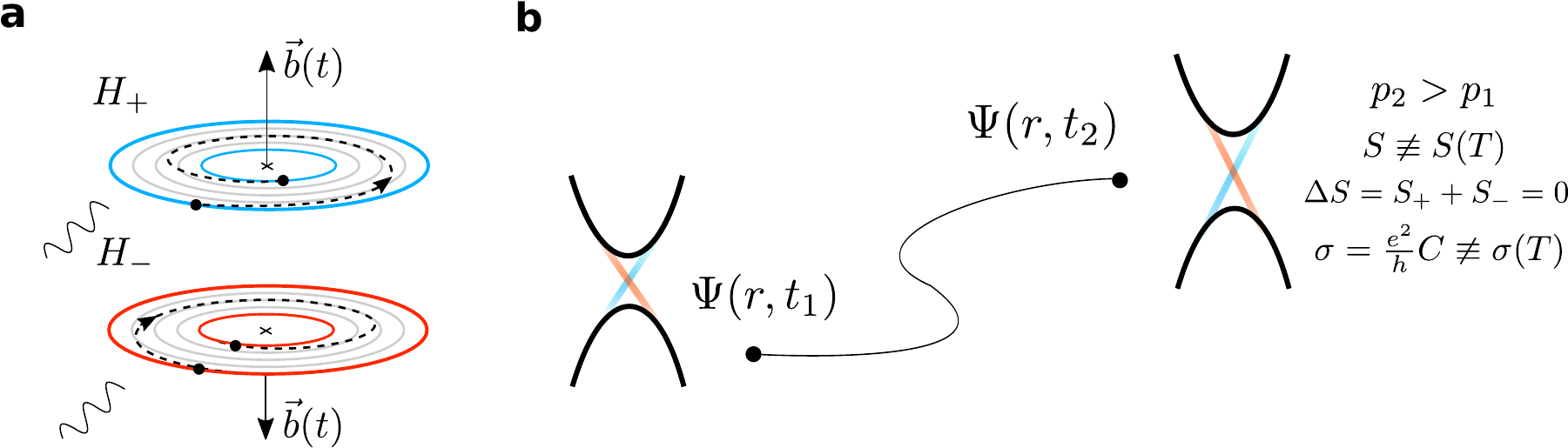}
    \caption{\textbf{Adiabatic coherent process} (\textbf{a}) Illustration of an adiabatic coherent process, for which the magnitude of the field $b$ increases. Blue a red circles represent the initial and resulting final states on each branch defined in $H_{2D}$. (\textbf{b}) Adiabatic coherent evolution of electronic wavefuntion $\Psi (r,t)$. Seebeck coefficient and entropy are not modified in this process allowing heat-electricity transformation in a completely reversible process. As a consequence of the enlargement of the $b$ field and the Berry curvature, the moment of the possible final states $p_2$ are higher than the initial electron moment $p_1$. However, in the limit in which oscillations does not change the Chern number, states remain to be ballistic with their conductance quantized and independent of temperature.}
\end{figure}

\clearpage
\pagestyle{empty}
\begin{center}
{\section*{{\normalfont Supplementary information for}\\ \vspace{12pt} Emergent topological fields and relativistic phonons within the thermoelectricity in topological insulator}

\large{\author
{Daniel Failde,$^{1\ast}$ Daniel Baldomir$^{1\ast}$\\} \vspace{12pt}
\normalsize{$^{1}$Departamento de Física Aplicada, Instituto de Investigacións Tecnolóxicas},\\
\normalsize{Universidade de Santiago de Compostela,}\\
\normalsize{E-15782 Campus Vida s/n, Santiago de Compostela, Spain}\\
\normalsize{daniel.failde.balea@rai.usc.es, daniel.baldomir@usc.es}
}}
\end{center}
\clearpage

\renewcommand{\thefigure}{\,S\arabic{figure}}
\setcounter{figure}{0}

Topological insulators (TIs) are Quantum Spin Hall (QSH) systems with conductor states on their surface associated with a non-trivial topology of their electronic band structure \cite{Konig766}. On their edge, bands follow a linear dispersion law $E=\hbar v_F k$ as it would happen for a relativistic particle with zero rest mass which goes through singularity points (Dirac points). These singularities are the sources of the non-trivial topology, which, protected by time-reversal symmetry induce chiral Kramers currents that can be determined using Berry's gauge fields associated with their curvature on U(1) or SU(N) groups depending on band degeneracy \cite{RevModPhys.82.3045,PhysRevB.78.195424}. The robustness of these spin-momentum locking channels as well as their quantized transport properties, make three and two-dimensional TIs strong candidates in the context of  Quantum Computing or Thermoelectric and Superconducting Devices. However, for a better understanding of these applications, some key points as thermal excitations and phonons have to be incorporated to electrons dynamic in TIs since their presence could lead to losing the quantum adiabaticity and coherence necessary to maintain such topological order \cite{PhysRevLett.107.210501}.Along this line, we analyze the possible effect of phonons and thermal excitations in TIs surface states by studying adiabatic mechanical oscillations through Dirac oscillator model \cite{moshinsky1989dirac}. In order to see if non-trivial topology is preserved under certain values for the involved physical magnitudes in this process, we make use of the field interpretation of the Berry curvature, where magnetic flux quantization of helical currents provides an argument to estimate the strength of the field $b$ associated to the topological regime. With this background, we can find an equivalence between the oscillation frequency and the intrinsic field at the same time we analytically verify the robustness of the topological regime against phonons, thermal gradients and similar perturbations. 

\newpage

\subsection*{Supplementary Note 1. The Dirac oscillator}
The introduction of mechanical oscillations in a relativistic context was first analyzed by M. Moshisnky and A. Szczepaniak \cite{moshinsky1989dirac} incorporating a linear term in $r$ to the Dirac equation. The origin of this term lies in the introduction of an harmonic oscillator potential into the Klein-Gordon equation, leading to the well-known Dirac oscillator
\begin{equation}
i\hbar (\partial \psi / \partial t)=[v_F \boldsymbol{\alpha}(\boldsymbol{p}-im \boldsymbol{r}\omega\beta)+mv_F^2\beta] \psi
\label{Dirac oscillator}
\end{equation}
being $\alpha_i=\left[\begin{array}{cc} {0} & \sigma_i \\
\sigma_i & 0 \\
\end{array} \right]$, $\beta=\left[\begin{array}{cc} \sigma_0 & 0 \\
0 & -\sigma_0 \\
\end{array} \right]$ , $\sigma_i$ the Pauli matrices, $m$ the mass of the particle, $r$ the position and where we have substituted the original speed of light $c$ by the Fermi velocity $v_F$ in order to adapt equation Eq.\,\ref{Dirac oscillator} into the context of TIs. Afterwards, the equation was further analyzed, always in the context of Quantum Field Theory (QFT) \cite{PhysRevA.76.041801,Rozmej_1999}, where  working with phonons it is always convenient to employ operators defined on a Fock space and Eq.\,\ref{Dirac oscillator} can be rewritten in function of the right and left chiral annhilation and creation operators $a_r = \frac{1}{ \sqrt{2}}(a_x -ia_y), a_r^+ = \frac{1}{\sqrt{2}}(a_x^+ +ia_y^+)$  and $a_l = \frac{1}{\sqrt{2}}(a_x +ia_y), a_l^+ = \frac{1}{\sqrt{2}}(a_x^+ -ia_y^+)$, being $a_x, a_y,a_x^+$ and $a_y^+$ the usual annihilation and creation operators of the harmonic oscillator. There are two Pauli spinor eigenstates which present entanglement between the spin and orbital degrees of freedom. 
\begin{equation}
    \ket{\psi_1}=i\frac{2mv_F^2\sqrt{\epsilon}}{E-mv_F^2}a_l^+\ket{\psi_2} 
\end{equation}

\begin{equation}
    \ket{\psi_2}=-i\frac{2mv_F^2\sqrt{\epsilon}}{E+mv_F^2}a_l^+\ket{\psi_1}
\end{equation}
being $\ket{\psi_1}$ and $\ket{\psi_2}$ the two components of the spinor $\ket{\psi}$, $\epsilon=\frac{\hbar \omega}{m v_F^2}$ takes into account the non-relativistic limit and $\ket{n_l}= \frac{1}{\sqrt{n_l!}}(a_l^+)^{n_l}\ket{0}$ the basis in which the Fock space is expanded \cite{PhysRevA.76.041801}. The energy spectrum is $E= \pm E_{n_l}= \pm m v_F^2 \sqrt{4 \epsilon{n_l +1}}$, whose eigenstates can be written as Pauli spinors $\ket{\phi_\uparrow}$ and $\ket{\phi_\downarrow}$ components, employing the angular momentum z-component definition given by  $L_z=\hbar (a_r^+a_r-a_l^+a_l)$

\begin{equation} 
    \ket{-E_{n_l}}=\beta_{n_l}\ket{n_l}\ket{\phi_\uparrow}+i\alpha_{n_l}\ket{n_l-1}\ket{\phi_\downarrow}
\end{equation}

\begin{equation}
    \ket{E_{n_l}}=\alpha_{n_l}\ket{n_l}\ket{\phi_\uparrow}-i\beta_{n_l}\ket{n_l-1}\ket{\phi_\downarrow}
\end{equation}
where $\alpha_{n_l}=\sqrt{\frac{E_{n_l}+mv_F^2}{2E_{n_l}}}$ and $\beta_{n_l}=\sqrt{\frac{E_{n_l}-mv_F^2}{2E_{n_l}}}$. Finally, time dependent state of the spinors excited by the Dirac oscillator is
\begin{equation}
    \ket{\psi(t)}=\left(cos\; \omega_{n_l}t+\frac{i}{\sqrt{4\epsilon n_l+1}}sin \; \omega_{n_l}t   \right) \ket{n_l-1} \ket{\phi_\uparrow} +\left(\sqrt{\frac{4\epsilon_{n_l}}{4\epsilon_{n_l}+1}} sin \; \omega_{n_l}t\right)\ket{n_l} \ket{\phi_\downarrow}
\end{equation}
Therefore, we see how there is one oscillation between the spin-orbit states $\ket{n_l-1} \ket{\phi_\uparrow}$ and $\ket{n_l} \ket{\phi_\downarrow}$, in such a form that the change of spin polarization implies one for the orbital and vice versa. However, these abstracts does not take into account the topology of the system and are not enough to treat relativistic phonons into TIs. With this purpose we are going to introduce it into the adiabatic context to see explicitly how they affect to the topological properties and thus to the thermoelectric response in TIs.

\subsection*{Supplementary Note 2. The adiabatic Dirac oscillator}

Coming back to the Eq.\,\ref{Dirac oscillator} we can rearrange its basis allowing its separation into two non-interacting and time-reversal counterparts $H_\pm (\boldsymbol{k}\mp e/\hbar \; \boldsymbol{A})$ in the same way as the 2D effective Hamiltonian used to describe the physics inside 2D and 3D TIs thin-films

\begin{equation}
H_\pm (\boldsymbol{k'})=\left[\begin{array}{cc} \pm m_e v_F^2 & {\hbar v_F[(k_x \mp e /\hbar \; A_x) -i (k_y \mp e/\hbar \; A_y)]} \\
{\hbar v_F[(k_x \mp e/\hbar \; A_x) +i (k_y \mp e/\hbar \; A_y)]} & \mp m_e v_F^2
\end{array} \right]
\label{Separating Dirac oscillator}
\end{equation}
where $\boldsymbol{A}=(-m\omega y/e, m\omega x/e,0)=(-\mathcal{B}y/2,\mathcal{B}x/2,0)$ is the vector potential which defines a magnetic field $\mathcal{B}$ of opposite sign for each branch of $H_\pm$. The form in which perturbation enters guaranties time-reversal symmetry conservation and its spin-orbit nature manifested in its non-relativistic limit. Given that $H\pm$ are non-interacting we can work with one of the subsystems (we choose $H_+$) and then extend our results to the other. The unperturbed eigenstates of Eq.\,\ref{Separating Dirac oscillator} are the eigenstates of the Dirac Hamiltonian $H_{2D}$, which for $H_+$ results
\begin{equation}
    \ket{n}=\frac{1}{\sqrt{2}}\left[\begin{array}{c} \sqrt{1+\frac{M({\bf k})}{\xi}} \\ e^{i\phi} \sqrt{1-\frac{M({\bf k})}{\xi}} \end{array}\right] \qquad \ket{m}=\frac{1}{\sqrt{2}}\left[\begin{array}{c} \sqrt{1-\frac{M({\bf k})}{\xi}} \\ -e^{i\phi} \sqrt{1+\frac{M({\bf k})}{\xi}} \end{array}\right]
\end{equation}
being $\ket{n}$ and $\ket{m}$ correspond to the states associated to the positive and negative energy solutions $\xi=\pm \sqrt{M^2(k)+\hbar^2 v_F^2 k^2}$. Generally speaking, $M(k)=M-Bk^2$, but this case will be discussed later given that our field $\boldsymbol{b}$ has been calculated without taking into account any dependence on the Hamiltonian parameter $B$. Thus, for the moment we are going to consider $M(k)=m_e v_F^2$. Now we are going to compute the correction to the eigenstates, that for an adiabatic perturbation and taking into account that $\xi_n-\xi_m=2\xi$ results to be

\begin{equation}
    \ket{n} \rightarrow  \ket{n} - i\hbar \frac{\partial k_x}{\partial t} \frac{\bra{m} \ket{\partial_{k_x} n}}{2 \xi} \ket{m} - i\hbar \frac{\partial k_y}{\partial t} \frac{\bra{m} \ket{\partial_{k_y} n}}{2 \xi} \ket{m}
\end{equation}
\begin{equation}
    \ket{m} \rightarrow \ket{m} + i\hbar \frac{\partial k_x}{\partial t} \frac{\bra{n} \ket{\partial_{k_x} m}}{2 \xi} \ket{n} + i\hbar \frac{\partial k_y}{\partial t} \frac{\bra{n} \ket{\partial_{k_y} m}}{2 \xi} \ket{n}
\end{equation}
It is the moment to introduce the perturbation given by the Dirac oscillator, which as we saw, enters into the form of a magnetic field for each branch. We have to be careful about the term $\partial k_i/ \partial t$ because quantically the velocity $\hat{v}$ and the momentum operator $\hat{p}$ present a different relationship to which present in a classical system. In relativistic quantum mechanics the velocity operator is defined as $\hat{v}_j=d \hat{x}_j/dt=i\hbar [\hat{H},\hat{x}]=c \hat{\alpha}_i$, so that it is more linked to the spin of the particle $S_j=\hbar/2 \alpha_j$ than to the moment. In contrast, we are introducing a Lorenz force in the system, in such a way that the correction to the momentum for a given direction must be proportional to the momentum in a perpendicular direction, that is why we take $\partial k_i/ \partial t=\epsilon_{ijk} k_j \omega_k$ in order to be consistent. Thus, we can proceed to rewrite the correction to the eigenstates

\begin{equation}
    \ket{n} \rightarrow  \ket{n} + \frac{\hbar \omega}{2\xi} \frac{\hbar v_F k}{2 \xi} \ket{m} \qquad \qquad \ket{m} \rightarrow  \ket{m} - \frac{\hbar \omega}{2\xi} \frac{\hbar v_F k}{2 \xi} \ket{n}
\end{equation}
and to compute analytically Berry curvature corrections $\Omega_{k_x,k_y}^n=i(\bra{\partial_{k_x}n}\ket{\partial_{k_y}n}-\bra{\partial_{k_y}n}\ket{\partial_{k_x}n})$ after some algebra. Several cases will be analyzed which demonstrates the correctness of the topological intrinsic $b$ that we have calculated.

\subsection*{Constant  $\omega$}

The first case corresponds to a purely constant $\omega$, i.e, the phonon frequency does not incorporate any dependence on the position $r$, the energy or the temperature. In that situation, the correction to the Berry curvature of the conduction band of $H_+$ results (Fig. S1)

\begin{equation}
    \Omega_{k_x,k_y}^n \rightarrow \Omega_{k_x,k_y}^n-\frac{\hbar\omega}{2\xi} \frac{M}{\xi} \Omega_{k_x,k_y}^n+\frac{\hbar\omega}{2\xi} \frac{\hbar^2v_F^2}{2\xi^2}-\frac{\hbar\omega}{2\xi} \frac{\hbar^2v_F^2}{\xi^2} \left(1-\frac{M^2}{\xi^2}\right)
\end{equation}
\begin{figure}
    \centering
    \includegraphics[scale=0.8]{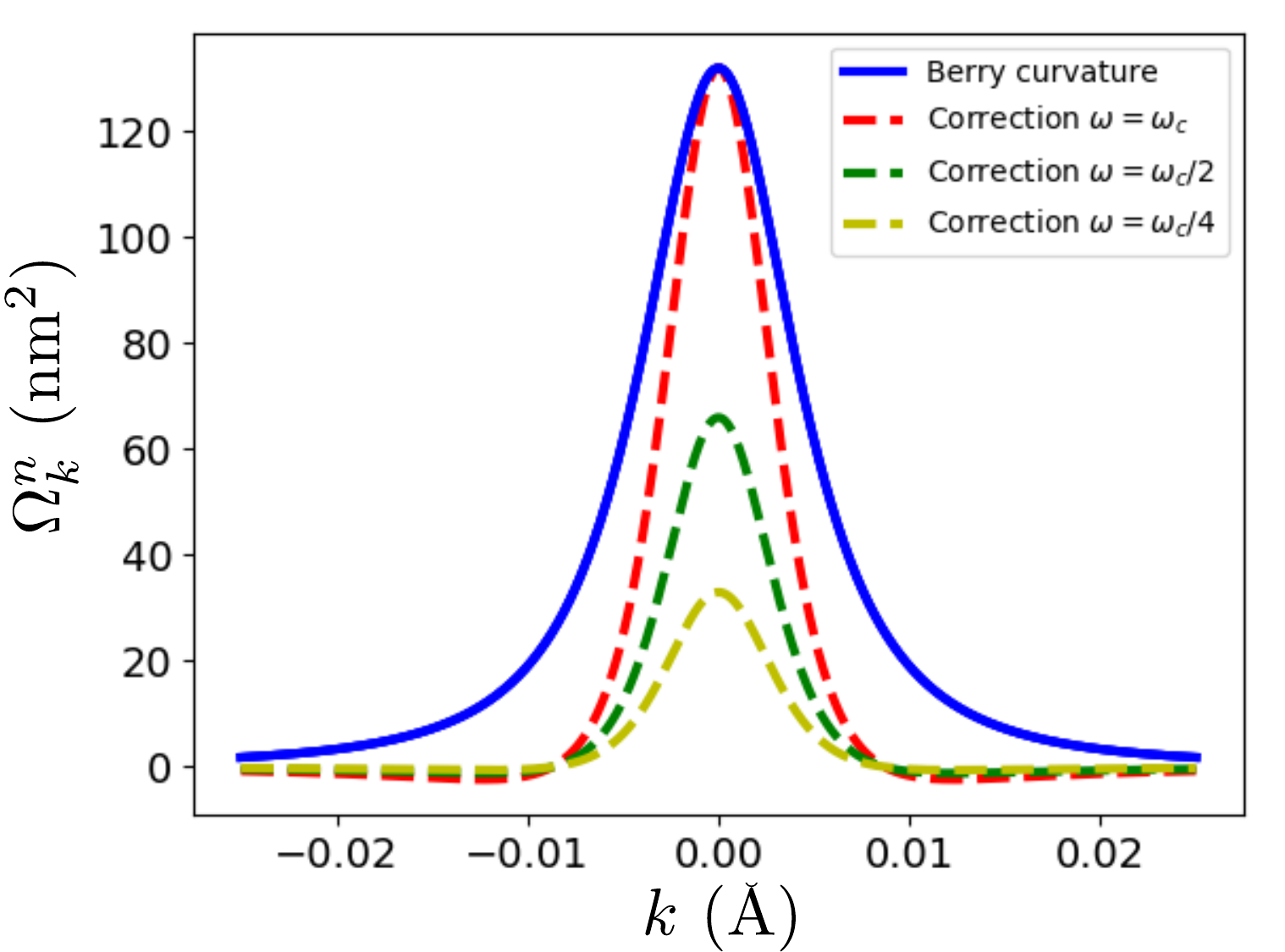}
    \caption{Unperturbed Berry curvature (blue solid line) of the conduction band of a Dirac Hamiltonian $H_+$ and first-order correction to it (dashed lines) for different constant frequencies below the critical value $\omega_c=eb/2m$. The parameters used are $M=-0.025$ eV, $B=0$ and $v_F=6.17$ $10^5$m/s.}
\end{figure}
The perturbation introduced produces a correction to the Berry curvature whose direction will depend on the sign of $\omega$. It is straightforward to note that the corrections introduced preserves particle-hole symmetry $\hat{C}$ ($\xi\rightarrow -\xi$) as well as time-reversal $\hat{T}$ ($M\rightarrow -M, \omega \rightarrow - \omega$). At critical frequencies $\omega_c=eb/2m$, the magnitude of the perturbation has the value of the unperturbed Berry curvature at $k=0$ evidencing the good interpretation of our approximation for the topological intrinsic field $b\approx 2M^2/\hbar e v_F^2$ as a measure of the robustness of the topological regime. The behaviour of the corrections is what one can expect from a constant perturbation into the real space and its homologous field into the $k$-space. This case might be enough to introduce certain phonons in topological insulators, however, it can not be applied generally given that thermal excitations and the majority of the phonons have an energy dependent nature.

\subsection*{Energy dependent $\omega$}
Given the physical equivalence between the frequency $\omega$ given by the Dirac oscillator Hamiltonian and the field $b=2m \xi/\hbar e$ we are going to suppose the phonon frequency to have an energy dependence of the type $\omega=eb/2m=\xi/\hbar$.

\begin{equation}
    \ket{n} \rightarrow  \ket{n} + \frac{\lambda}{2} \frac{\hbar v_F k}{2 \xi} \ket{m} \qquad \qquad \ket{m} \rightarrow  \ket{m} - \frac{\lambda}{2} \frac{\hbar v_F k}{2 \xi} \ket{n}
\end{equation}
This allows us exploring the interpretation on the $b$ field and to demonstrate that oscillation below this limit could modulate the Berry curvature by non producing changes into their topological properties. If our interpretation of the field $b$ as a translation of the Berry curvature onto the real space is correct, then its corrections should result in a function of the type of the unperturbed Berry curvature. In this case, the correction to the Berry curvature is slightly different from the previous situation

\begin{equation}
    \Omega_{k_x,k_y}^n \rightarrow \Omega_{k_x,k_y}^n+ \lambda \frac{1}{2}\left(-\frac{\hbar^2v_F^2}{2\xi^2}+\frac{\hbar^2v_F^2}{2\xi^2}\frac{\hbar^2v_F^2k^2}{\xi^2}-\frac{M}{\xi}\Omega^n\right)=\Omega^n- \lambda\frac{M}{\xi}\Omega^n
\end{equation}
where $\lambda$ is a dimensionless parameter to measure the strength $[0,1]$ of the perturbation, i.e, $\lambda=1$ correspond to the case of a frequency $\omega=eb/2m$. The correction results in a simpler expression and which fits better to a function of the type of the Berry curvature. Although it is masked in the calculation, since we cancelled both $\omega$ and $\xi$, one can also check that the results also preserves $\hat{C}$ and $\hat{T}$ symmetries as in the previous case.  With these ingredients, we are in a position to treat phonons and thermal oscillations into TIs. Despite of the conditions seem to be a bit restrictive, given that we are considering only in-plane oscillations, they are not unrealistic. It has been shown how polar optical modes can strongly couple with the topological electrons when the Fermi level lies close to the Dirac point as well as it has been reported how these modes present a linear dispersion law at low $k$ that allow them to fulfill the conditions underlined in the article \cite{Heid2017}.

\subsection*{Incorporating the Hamiltonian parameter B}
One more case it is needed to complete the current analysis and it corresponds to incorporate the $-Bk^2$ term to the particles mass $M(k)$. This should be done to put the previous calculations closer to the context of TIs rather than in purely relativistic quantum mechanics in the vacuum. Obviously, in part, it has been already done once we considered the Fermi velocity $v_F$ instead of the light velocity $c$ in our formalism. Essentially, corrections remain equal with the substitution of $M$ by $M-Bk^2$ but we have to add an additional term in the case of a constant $\omega$. Thus, in the first case

\begin{equation}
    \Omega_{k_x,k_y}^n \rightarrow \Omega_{k_x,k_y}^n-\frac{\hbar\omega}{2\xi} \frac{M(k)}{\xi} \Omega_{k_x,k_y}^n+\frac{\hbar\omega}{2\xi} \frac{\hbar^2v_F^2}{2\xi^2}-\frac{\hbar\omega}{2\xi} \frac{\hbar^2v_F^2}{\xi^2} \frac{\hbar^2v_F^2k^2}{\xi^2}+\frac{\hbar\omega}{2\xi} \frac{\hbar^2v_F^2}{\xi^2} \frac{M(k)2Bk^2}{\xi^2}
\end{equation}
where here clearly $\Omega^n=-\frac{\hbar^2v_F^2 (M+Bk^2)}{2\xi^3}$ and $M(k)=M-Bk^2$. For an energy dependent $\omega$
\begin{equation}
    \Omega^n \rightarrow \Omega^n- \lambda\frac{(M-Bk^2)}{\xi}\Omega^n
\end{equation}
Due to the addition of the $-Bk^2$ term corrections show to change their sign at a certain $k$ in contrast with the Berry curvature. This is easily justified given that our approximation for $b$ and thus for $\omega$ does not take into account any dependence on this term. Despite this, given that our approximation is valid for systems satisfying  $v_F^2>>2MB/\hbar^2$, which is easily fulfilled thanks to the small gap and high Fermi velocity that typically characterizes 3DTI thin films, the addition of this term to the unperturbed Berry curvature does not produce any change of its sign in the mentioned limits where corrections tend to zero faster than $\Omega^n$.
\begin{figure}
    \centering
    \includegraphics[scale=0.6]{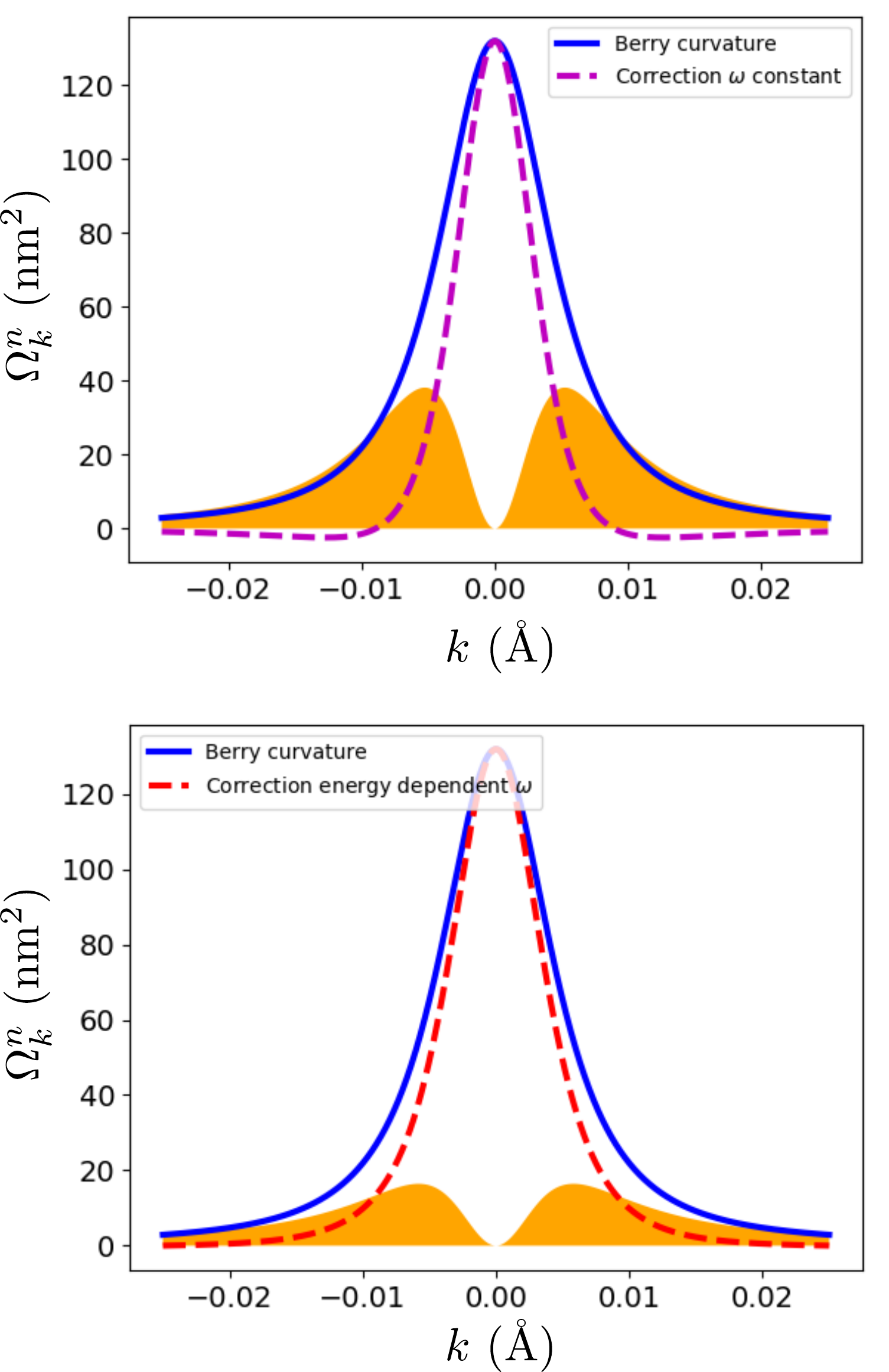}
    \caption{Unperturbed Berry curvature (blue solid line) of the conduction band of a non-trivial Dirac Hamiltonian $H_+$ ($M<0, B<0$) and first-order correction to it (magenta and red dashed lines) for critical frequencies $\omega=eb/2m$ considering constant and energy dependences respectively. The parameters used are $M=-0.025$ eV, $B=-20$ eV\AA$^2$ and $v_F=6.17$ $10^5$m/s. Filled regions correspond to the difference between the unperturbed Berry curvature and the first-order corrections showing no sign inversion even at critical frequencies.}
\end{figure}
\end{document}